\documentclass[twocolumn,showpacs,pra,superscriptaddress]{revtex4} 
\usepackage[final]{graphicx}
\usepackage{mathrsfs}
\usepackage{amssymb}

\begin{document}

\title{Experimental investigation of left-right asymmetry in photon-atom interaction}

\author{S. Ricz}
\affiliation{Institute for Nuclear Research, Hungarian Academy of
Sciences, Debrecen, P.O. Box 51, H-4001, Hungary}
\affiliation{Institute for Atomic and Molecular Physics,
Justus-Liebig University Giessen, D-35392 Giessen, Germany}
\author{T. Buhr}
\affiliation {Physikalisch-Technische Bundesanstalt, Bundesallee 100, D-38116 Braunschweig, Germany}
\author{\'A. K\"ov\'er}
\affiliation{Institute for Nuclear Research, Hungarian Academy of Science, Debrecen, P.O. Box 51, H-4001, Hungary}
\author{K. Holste}
\author{A. Borovik, Jr.}
\author{S. Schippers}
\affiliation{Institute for Atomic and Molecular Physics, Justus-Liebig University Giessen, D-35392 Giessen, Germany}
\author{D. Varga}
\affiliation{Institute for Nuclear Research, Hungarian Academy of Science, Debrecen, P.O. Box 51, H-4001, Hungary}
\author{A. M\"uller}
\affiliation{Institute for Atomic and Molecular Physics, Justus-Liebig University Giessen, D-35392 Giessen, Germany}

\date{\today}


\begin{abstract}
Single ionization of noble gas atoms by linearly polarized
synchrotron radiation has been studied by employing angle- and
energy-resolved photoelectron spectroscopy. The measurements were carried out in the plane
  defined by the momentum and polarization vectors of the photon.
  Parameters describing the left-right asymmetry (LRA)
(relative to the photon propagation direction) of the photoelectron
angular distribution were determined experimentally for the
$s$-shells  of He, Ne, Ar, Kr and Xe atoms and H$_2$ molecules and
for the $p$-shells of Ne, Ar, Kr and Xe atoms. The values of the
left-right asymmetry
 differ significantly from zero for both subshells. The photon and photoelectron energy dependence of the
LRA parameters are presented also. Possible experimental and
instrumental sources that could generate asymmetry are discussed and
excluded as well.
\end{abstract}

\maketitle

\section{Introduction}

According to quantum mechanics electromagnetic atomic transitions
have space inversion symmetry. The electromagnetic interactions
among the atomic electrons and nucleus as well as between the
ionized and excited particles are assumed to conserve parity.
Consequently for photoionization with linearly polarized light
 the angular distributions of the emitted particles or
quanta should show left-right symmetry relative to the photon
propagation direction. In other words, the right
side intensity of ejected particles with respect to the beam
direction is equal to the left side intensity.

\begin{figure}
\centering
\includegraphics[width=8.5cm]{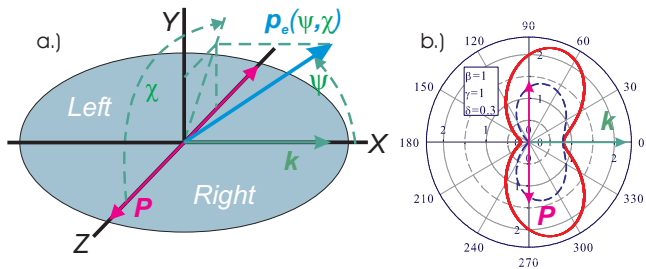}
\caption{(Color online) (a) definition of the coordinate system. (b)
an example for the double differential cross section of
photoelectrons in the scattering
 ({\bf P,k}) plane. The solid and the dashed lines represent the
angular distribution of emitted photoelectrons for linearly
polarized and unpolarized light calculated from Eq.
\ref{eqn:equ_unpol} and Eq. \ref{eqn:equ_pol}, respectively for the
given parameter values.} \label{fig:geom}
\end{figure}

The angular distribution of photoelectrons can be expressed by the
following formulae for unpolarized (\hbox{Eq. \ref{eqn:equ_unpol}})
and entirely ($100\%$) linearly polarized (Eq. \ref{eqn:equ_pol})
light including first-order
 non-dipole correction (\cite{Cooper1990,Derevianko1999}):

\begin{widetext}
\begin{equation}
\label{eqn:equ_unpol} \frac{d\sigma _{nl}}{d\Omega} =\frac{\sigma
_{nl}}{4\pi \Large}[1-\frac{1}{2}\beta P_2(\cos \psi)+(\delta
+\frac{1}{2}\gamma\sin ^2 \psi)\cos\psi \Large]
\end{equation}

\begin{equation}
\label{eqn:equ_pol} \frac{d\sigma_{nl}}{d\Omega} =\frac{\sigma
_{nl}}{4\pi} \Large[1+\frac{1}{2}\beta (3\sin ^2 \psi cos ^2 \chi
-1) + (\delta +\gamma\\sin ^2 \psi cos ^2 \chi) \cos\psi \Large],
\end{equation}

\end{widetext}

\noindent where $\beta$ is the dipole (E1), $\delta$ and $\gamma$
are the non-dipole (M1, E2) anisotropy parameters. The definition of
the angles is shown
 in figure \ref{fig:geom}\ a.  The polar ($\psi$) and the
azimuthal ($\chi$) angles of the photoelectrons are measured with
respect to the momentum (${\bf k}$) and polarization vectors (${\bf
P}$) of the photons,
 respectively (figure\ \ref{fig:geom}\ a). $P_2$ is the
second order Legendre polynomial, $\sigma_{nl}$ is the
photoionization cross section for the $nl$ shell. An example
 for the photoelectron angular distribution is shown in figure\ \ref {fig:geom}\ b.
As it is seen the non-dipole correction produces only a
forward-backward asymmetry relative to the photon momentum vector in
the case of linearly polarized and unpolarized photon beams.
Therefore, it does not break the symmetry around the propagation
direction of the photon. For a linearly polarized photon beam a
right-handed XYZ coordinate system (figure\ \ref{fig:geom}\ a) can
be defined in the following way \cite{Ricz2007}: The photon momentum
vector $\vec k$ points to the direction of the X-axis and the photon
polarization vector ($\vec P$) is aligned along the Z-axis. The
Y-axis is perpendicular to the XZ-plane and points upward. (The
$\vec E(r, t)$ electric vector of the incoming photon oscillates in
the XZ-plane.) Positive Z defines the right hand side (R) and
negative Z the left hand side (L) relative to the propagation
direction of the incoming light. The mirror plane (XY-plane in
igure\ref{fig:geom}a) is perpendicular to the polarization direction
of the photon beam. Using the above described coordinate system the
left-right asymmetry parameter $A_{LR}$ for photoelectron emission
may be defined as \cite{Ricz2007} :

\begin{eqnarray}
\label{eqn:asym}
{A_{LR}=\frac{\sigma_L-\sigma_R}{\sigma_L+\sigma_R}},
\end{eqnarray}

\noindent where $\sigma_L$ and $\sigma_R$  are the cross sections for
photoelectron emission to the left (L) side
($\psi$  varies between $0^{\circ}$ and $180^{\circ}$, clockwise) and to the right side
(R) ($\psi$ varies between $0^{\circ}$ and $-180^{\circ}$, counterclockwise)
relative to the photon propagation direction, respectively.

In our previous paper \cite{Ricz2007} left-right asymmetry (LRA)
has been observed in the outer $s$-shell double differential photoionization
cross sections for linearly polarized synchrotron radiation.
The investigated angular ranges for photoelectron emission were
 $\chi = \pm 1.7^{\circ}$ and \hbox{$\psi = 0^o \pm180^{\circ}$}.
 For the interpretation of the experimental data two
possible explanations for the observed LRA were suggested and
excluded: (a) The LRA is the result of the weak interaction among
the nucleons and the atomic electrons mediated by the exchange of
$Z_0$ bosons. However, the experimental atomic mass dependence and
the order of magnitude of the experimental values strongly differ
from the theoretical predictions published in \cite{Ricz2007}. This
indicates that the observed left-right asymmetry cannot be
interpreted as a result of the weak interaction. (b) The LRA is
associated with the interactions of ultra short laser pulses
\cite{Paulus}. When the photon wave packet is extremely short and
the phase difference of the carrier-envelope is constant, LRA (or
virtual parity violation) may exist. However, such time structure of
photon wave packets emitted by a synchrotron light source has not
been observed so far.

The aim of the present work was to check the previously observed
left-right asymmetry \cite{Ricz2007} by applying different
experimental setup (electron spectrometer and synchrotron). The
measurements was carried out at the DORIS\ III synchrotron light
source \cite{Reininger,Castro,Moller, Larsson},
 while the previous experiments \cite{Ricz2007}
 were done at the MAX\ II synchrotron \cite{Bassler1,Bassler2}.
 The preliminary results were published
by Rics\'oka {\it et al} \cite{Ricsoka2009}. Furthermore, our
experimental investigation has been extended to the $p$-shells of
Ne, Ar, Kr and Xe using nearly the same photon energies as for the
corresponding $s$-shells. Moreover, the previous spectra measured at
MAX\ II synchrotron were also reevaluated for the determination of
the LRA parameters.
 In the Appendix the possible experimental and
instrumental effects, which may cause LRA and their magnitudes are
carefully analyzed.

\section{Experiment and data evaluation}

Two different synchrotrons were used in the measurements: the DORIS\
III storage ring at HASYLAB, Hamburg, Germany (beam line BW3) and
the MAX\ II at MAX-lab, Lund, Sweden (beam line I411).
 Positrons are used to produce synchrotron radiation in DORIS\ III and
 the operating energy is 4.45 GeV. This energy is almost three times higher
than that of MAX\ II (1.5 GeV) where electrons are accelerated. The
photon source at beam line BW3 was composed of two undulators with
overlapping energy ranges, while at beam line I411 of the MAX\ II
synchrotron one undulator is used. Both beam lines were equipped
with SX-700 monochromators.

\begin{figure}[h]
\centering
\includegraphics[width=9cm]{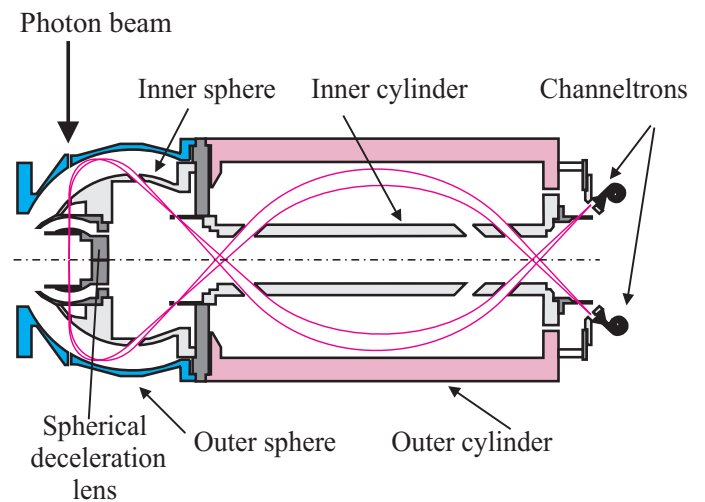}
\caption{(Color online) Schematic cross section of an ESA-22 type
electron spectrometer. \label{fig:spm}}
\end{figure}

The ejected photoelectrons were analyzed with
the ESA-22G and ESA-22L \cite{ricz2002} electrostatic electron spectrometers.
The ESA-22G is a slightly modified version of the ESA-22L electron spectrometer
 developed in Atomki, Debrecen, Hungary. The main working principles and
geometric dimensions are the same for both spectrometers.
 A sketch of the analyzer can be seen in figure\ \ref{fig:spm}.
It consists of a spherical and a cylindrical electrostatic
mirror. The spherical part focuses the electrons from the scattering
plane (XZ-plane) to the entrance slit of the cylindrical mirror analyzer which
performs the energy analysis of the electrons.

\noindent A spherical deceleration lens is placed around the source
volume to improve the energy resolution of the equipment. Due to the
radial electrostatic field of the decelerator lens and of the whole
 spectrometer the polar angles of the emitted electrons
($\psi$ in figure\ \ref{fig:geom}) are conserved from the target all
the way to the detectors. The analyzer and the  interaction region
are shielded from the Earth's magnetic field by  three layers of
$\mu$-metal sheets. The residual magnetic field in the scattering
plane and in the analyzer is less than $500$ nT for both electron
spectrometers.

The photoelectrons were detected at $\chi=0^{\circ}$ azimuthal and
at 22 polar angles ($\psi$) in $15^{\circ}$ steps between
$0^{\circ}$ and $\pm180^{\circ}$ relative to the photon momentum
vector (figure\ \ref{fig:geom}). The acceptance angles of each
channeltrons were $\Delta\psi=\pm5^{\circ}$ and
$\Delta\chi=\pm1.7^{\circ}$. An important difference between the two
analyzers is the availability of two additional observation angles
in ESA-22G (at $90^{\circ}$ and $-90^{\circ}$ with respect to the
photon momentum vector) in the direction of the oscillating electric
field vector.
 Another difference between the two
spectrometers is the modified geometry of the gas target. (The gas
nozzle locates in the symmetry axes of both analyzer.) In case of
the ESA-22G a simple tube is used as a gas nozzle and the gas flows
upwards while in the ESA-22L \cite{ricz2002} a channelplate is fixed
to the end of the nozzle and the gas is directed downwards. In the
latter case a more directed gas flow is expected. In the DORIS\ III
experiment a new analyzer control and a faster signal processing
system as well as a new software were used to control the
spectrometer and to collect data.

In the recent experiments at DORIS\ III the LRA parameters were
determined at 203.3 eV photoelectron energy in the XZ-plane (it is
the same plane as in the previous experiment \cite{Ricz2007}). The source size was
$\pm 0.85$~mm determined by the geometry of ESA-22G in the direction perpendicular
 to the XZ-plane. The photon energies
were chosen such that the kinetic energies of the photoelectrons,
ejected from the outer $s$-shells, were nearly the same as those of
the Ar LMM Auger electrons. (The corresponding photon energy range
was \hbox{226.7\ -\ 256.3\ eV}.) In this way identical experimental
conditions were ensured for the detection of Auger- and
photoelectrons. The energy and angular distributions of the Auger-
and photoelectrons were measured at $80$\ eV pass energy (and the
energy resolution of the analyzer was about \hbox{$160$\ meV} full
width at half maximum (FWHM). For He $1s$ and Ne $2s$ shells a
\hbox{500 $\mu$m} wide monochromator slit size (corresponding to a
bandwidth of 40~meV) and for Ar $3s$ and Xe $5s$ shells a slit size
of 180 $\mu$m (corresponding to a bandwidth of 13~meV) were used.
This allowed us (together with the high energy resolution of the
spectrometer) to separate the satellite photolines from the diagram
lines.

\begin{figure}[h]
\centering
\includegraphics[width=9cm]{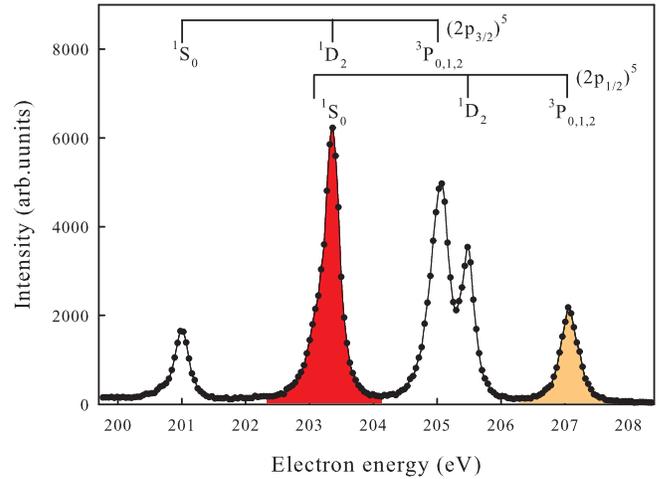}
\caption{(Color online) The Ar $L_{2,3}-M_{2,3}M_{2,3}$ Auger group
measured at 461.2 eV photon energy with a 500 $\mu$m monochromator
slit size. The colored Auger peaks were used to normalize the
photoelectron spectra. The dark red shaded peak
($\approx$\hbox{203.3\ eV})  was employed at the DORIS\ III
 experiment, while in the \hbox{MAX\ II} experiment \cite{Ricz2007} the light
orange line ($\approx$\hbox{207.0\ eV}) was used.
 \label{ArLMM}}
\end{figure}

We have not found any indication in the literature for nonzero asymmetry parameters
in the case of the Auger process for atoms. Therefore, the intensity
of all photoelectron spectra measured at different angular
channels were normalized to the intensity of the Ar LMM Auger peaks.
This normalization was necessary since the individual detection efficiencies
of the channeltrons were not known. In the previous experiments \cite{Ricz2007}
 the isotropic but weak Ar $L_2-M_{2,3}M_{2,3}$ $^3P_{0,1,2}$  Auger transitions
(at \hbox{207\ eV}, light orange peak in figure.\ \ref{ArLMM}) were
used to normalize the photoelectron spectra. In this measurement the
almost isotropic Ar $L_3-M_{2,3}M_{2,3}$ $^1D_2$ diagram Auger line
at 203.3 eV kinetic energy was used for normalization (see figure\
\ref{ArLMM}, dark red peak). It was measured at 461.2 eV incident
photons using 500 $\mu$m monochromator slit size. This line was
chosen because its intensity is nearly three times higher than that
of the isotropic Ar $L_2-M_{2,3}M_{2,3}$ $^3P_{0,1,2}$ Auger
transition. Its anisotropy is weak according to theoretical
calculations.

The angular distribution of Auger electrons resulting from the decay
of a $np_{3/2}$ vacancy (with principal quantum number
$n=2,3,4,...$) is given  as \cite{Cleff74, Berezhko77}:

\begin{eqnarray}
\label{Auger ang.distr.}
 {\frac{d\sigma(\theta)}{d\Omega}=\frac{\sigma_0}{4\pi}(1+\beta_2P_2(cos\theta))},
\end{eqnarray}

\noindent where $\frac{d\sigma(\theta)}{d\Omega}$ is the double
differential cross section of the Auger electron production at angle
$\theta$ with respect to the photon beam direction, $\Omega$ is the
solid angle, $\sigma_0$ is the total cross section, $\beta_2$ is the
anisotropy parameter and $P_2(cos\theta)$ is the second-order
Legendre polynomial. $\beta_2$ is related to the alignment parameter
($\mathscr{A}_{2}$) in the following way:

\begin{eqnarray}
\label{anisot par}
 {\beta_2=\alpha_2\mathscr{A}_{2}},
\end{eqnarray}

\noindent where $\alpha_2$ is the anisotropy coefficient. The
alignment parameter depends only on the ionization process, while
the anisotropy coefficient depends only on the decay process
(two-step model). In the case of the Ar \hbox{$L_2-M_{2,3}M_{2,3}$
$^3P_{0,1,2}$} transitions there is no alignment because the $J_i$
total angular momentum of the initial state equals $1/2$
($\mathscr{A}_{2}=0$) \cite{Cleff74, Mehlhorn68}. This means that
the angular distributions of these Auger electrons are isotropic.
The anisotropy parameter $\beta_2$ of \hbox{$L_3-M_{2,3}M_{2,3}$
$^1D_2$} transition calculated with Eq.\ \ref{anisot par} is
approximately -0.0288 (using the data in Ref.
\cite{Berezhko77,Berezhko78} for $\mathscr{A}_{2}$ and  in Ref.
\cite{Sarkadi90} for $\alpha_2$). Due to this small value the above
mentioned Auger transition can be used for normalization. Since the
angular distribution of
 the Auger electrons is
symmetric with respect to the photon beam momentum vector (as Eq.\
\ref{Auger ang.distr.} shows), the normalization of the
photoelectron angular distribution to an anisotropic angular
distribution of Auger electrons cannot introduce any LRA.

The linear polarization of the photon beam was monitored by
recording the angular distribution of Ne $2s$ photoelectrons at 250
eV photon energy where the non-dipole contribution is negligible
\cite{Hemmers}. The radiation was found to be completely linearly
polarized: $100$~\% within $2$~\% uncertainty.

\begin{figure}[ht]
\centering
\includegraphics[width=9cm]{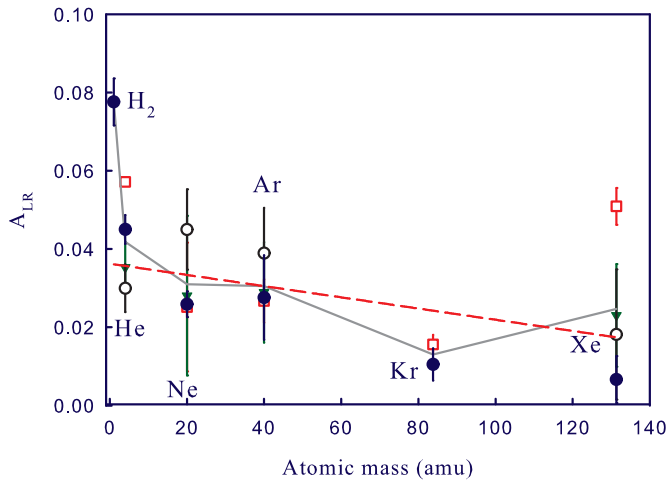}
\caption{(Color online) Comparison of the experimental LRA
parameters ($A_{LR}$) for H$_2$ molecule and outer $s$-shells of
noble gases. Three data sets were collected at MAX\ II
\cite{Ricz2007}  and one was measured at DORIS\ III
 \cite{Ricsoka2009}: $\bullet$ first observation of the
nonzero asymmetry parameters at MAX\ II; $\square$ results of
repeated experiment after the careful test of the experimental
system at MAX\ II; $\blacktriangledown$ asymmetry parameters after
rotation of the spectrometer system by $180^\circ$ horizontally at
MAX\ II;
 $\circ$ data measured at DORIS\ III. The solid line shows the average LRA
parameters obtained from the results of different measurements. The
dashed line is the linear fit for the mean values of the noble
gases.} \label{fig:mass}
\end{figure}

The measurement and the evaluation of the experimental data were
performed in a manner similar to the one described in Ref.\ \cite{Ricz2007}. The
photon flux was measured by a photodiode. The collection times of
the photoelectrons were several tens of seconds at each energy
point and the energy sweeps were repeated $10$ to $90$ times
depending on the magnitude of the photoionization cross sections.
Before and after the collection of photoelectron spectra an Ar LMM
Auger spectrum was recorded. After linear background substraction the
angular distribution of the photoelectrons was determined by
normalizing the intensity of the photoelectron line at each angle to
 the area of the selected (nearly isotropic) Auger peak in every single angular
channel. The relative double differential cross sections obtained at
different emission angles were summed for the left and
right spectrometers halves, separately.  Finally, the asymmetry
parameter was calculated by using Eq.\ \ref{eqn:asym} and its error
was estimated from the statistical uncertainty, background
substraction, normalization and reproducibility.

Possible experimental or instrumental sources that could generate
left-right asymmetry are discussed and their magnitudes are
estimated in the Appendix.

\section{Results and discussion}

Figure\ \ref{fig:mass} compares the experimental values of the LRA
parameters ($A_{LR}$) determined for the $s$-shell photoelectrons
for the H$_2$ molecule and noble gas atoms as a function of the
atomic mass. The figure shows four measured data sets. Three of them
were collected at \hbox{MAX\ II} synchrotron and published earlier
\cite{Ricz2007}. One of them was measured when the spectrometer
system was rotated by 180$^\circ$ horizontally around the analyzer
axis ($\blacktriangledown$). The fourth data set ($\circ$) was
measured at  DORIS III storage ring with the ESA-22G electron
spectrometer system \cite{Ricsoka2009}. All data sets show nearly
the same asymmetry. It is hard to believe that the time structure
and the phase differences between the carrier-envelope of the
photons are the same for both synchrotrons.
 This indicates that the asymmetric photoelectron emission
 is not an instrumental or experimental effect and originates
from the photoionization process itself. All four data sets show
definite positive asymmetry parameters for photoionization of the
$s$-shells of the studied targets.

\begin{figure}
\centering
\includegraphics[width=9cm]{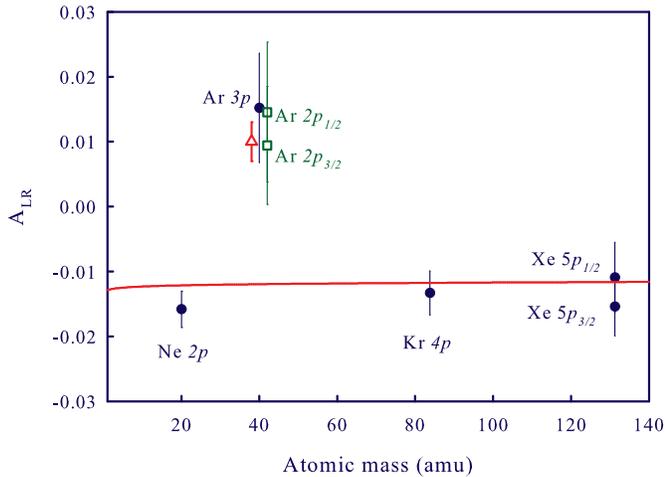}
\caption{(Color online) The experimental values of the asymmetry
parameters ($A_{LR}$) for the $p$-shells of Ne, Ar, Kr and Xe atoms.
\hbox{$\bullet$ present work;} $\square$ previously measured data
for Ar $2p$ shells \cite{Ricz2007}; $\bigtriangleup$ Ar $2p$ data
measured by Heinasmaki {\it et al} \cite{Heinasmaki2013}.}
\label{pshell}
\end{figure}

The LRA was also investigated for the Ne $2p$, Ar $3p$, Kr $4p$ and
Xe $5p$ shells
 at the MAX\ II synchrotron using the same experimental conditions as
 for the $s$-shells. Figure\ \ref{pshell} shows the asymmetry parameters as a function of the
atomic mass (Ar $2p$ asymmetry parameters were added to the figure
from \cite{Ricz2007, Heinasmaki2013}). As it is seen nearly constant
negative values were determined for the asymmetry parameters of Ne,
Kr and Xe atoms. Compared to the data measured for $s$-shells there
are significant differences. The sign of the values are opposite.
Considering only the atomic targets the absolute values of LRA are
higher for the $s$-shells than for the $p$-shells. (The mean
asymmetry parameter is $A_{LR}^s=0.028(5)$ for the $s$-shell without
$H_2$ and for the $p$-shell is $A_{LR}^p=-0.014(1)$ without Ar. The
standard errors are in the brackets.) The LRA data for $s$-shells
are decreasing slightly with increasing atomic mass (see dashed line
on \hbox {figure \ref{fig:mass}}) and the slope and intercept of the
linear fit are $-1.4(7)$x$10^{-4}$ and $0.036(5)$, respectively.
(The $H_2$ data was not taken into account in the fit.) The
experimental values are nearly constant for the $p$-shells (except
for Ar). These differences (see figures\ \ref{fig:mass} and
\ref{pshell}) indicate that the unknown correlation between the
photon and atom is sensitive to the angular momentum of atomic
shells. This shows that the asymmetry is an intrinsic behavior of a
photon-atom/molecule interaction.

\begin{figure}[ht]
\centering
\includegraphics[width=8.5cm]{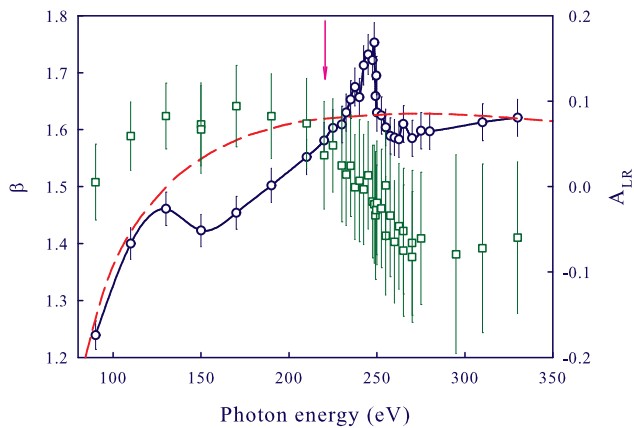}
\caption{(Color online) Dipole anisotropy ($\beta$) and LRA
parameters ($A_{LR}$) for the angular distribution of $3p$
photoelectrons of Ar as a function of the photon energy. $\circ$
experimental $\beta$ anisotropy parameters from \cite{Ricz2005}.
 The line is to guide the eye.
$\square$ experimental LRA parameters. The dashed red line shows the
theoretical calculation for the anisotropy $\beta$ parameter by
Derevianko {\it et al} \cite{Derevianko1999}. The arrow denotes the
photon energy where the asymmetry parameter was presented in figure\
\ref{pshell}.} \label{ar3pbeta}
\end{figure}

In contrast to the outer $p$-shell data the values of the asymmetry
parameter for Ar $2p$ and $3p$ shells are positive (figure\
\ref{pshell}). For further investigation of this behavior figure\
\ref{ar3pbeta} shows the dipole anisotropy and asymmetry parameters
for $3p$ photoelectrons of Ar as a function of the photon energy.
The deviation between the measured \cite{Ricz2005} and calculated
\cite{Derevianko1999} anisotropy parameters $\beta$ starts at
\hbox{$125$ eV} photon energy which is far from the $2p$
photoexcitation/photoionization threshold. It is an indication for
the strong correlation between the $3p$ and $2p$ shells of Ar. This
may be the reason while the LRA parameters for $3p$ and $2p$
photoelectrons are positive and roughly the same.

\begin{figure}
\centering
\includegraphics[width=9cm]{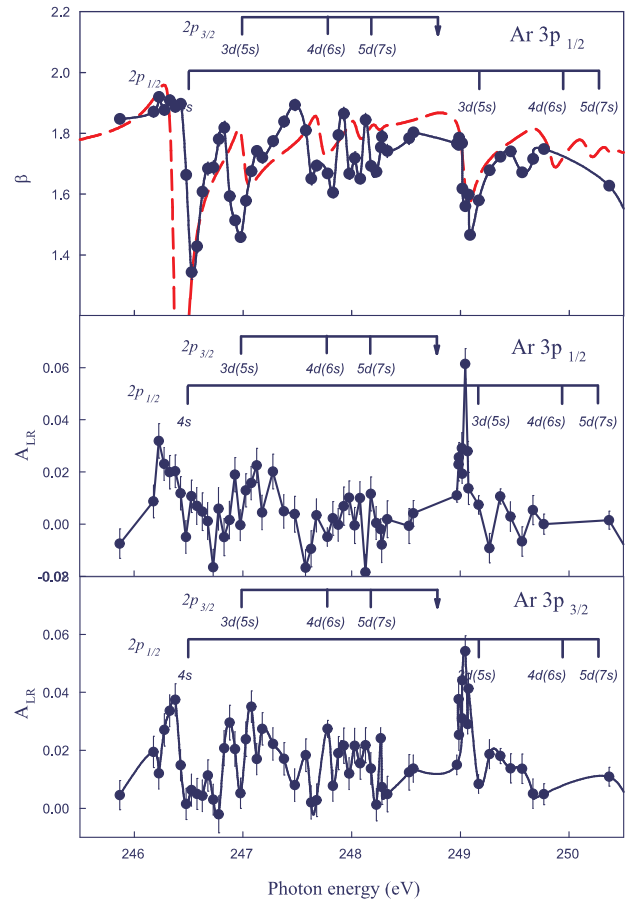}
\caption{(Color online) Dipole ($\beta$) (top panel for Ar
$3p_{1/2}$)
 anisotropy \cite{Ricz2005} and LRA ($A_{LR}$ ) parameters of Ar
$3p_{1/2}$ (middle panel) and $3p_{3/2}$ shells (bottom panel) in
the photon energy range where the resonant excitations of Ar
$2p$-shell exist. The vertical lines denote the positions of the
resonances. The dashed line in the top panel is the theoretical
calculation of Gorczyca {\it et al} \cite{Gorczyca}. }
\label{fig:asym-reso}
\end{figure}

In our previous paper \cite{Ricz2005} a detailed investigation for
the Ar $3p$ photoelectron angular distribution was carried out. The
angular anisotropy parameters for the \hbox{$(2p)^{-1}$-$ns/md$}
resonant photon energy region was determined with high energy
resolution and with narrow photon bandwidth. Here the LRA parameters
for the Ar $3p_{1/2}$  and $3p_{3/2}$ photoelectron angular
distributions are presented as a function of the photon energy
(figure\ \ref{fig:asym-reso}). Both the angular anisotropy ($\beta$)
and LRA parameters ($A_{LR}$) vary strongly in the vicinity of
\hbox{$(2p)^{-1}$-$ns/md$} resonances. These variations are
extremely sharp around the {$3d(5s)$} resonance. We think that such
sharp resonances cannot be produced by any experimental error.

\begin{figure}[ht]
\centering
\includegraphics[width=9cm]{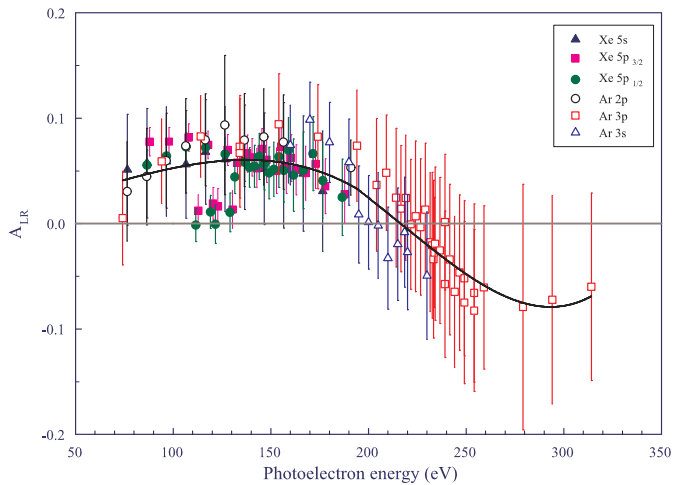}
\caption{(Color online) The photoelectron energy dependence of the
experimental LRA parameter ($A_{LR}$) in the cases of $s$- and
$p$-shells for Ar and Xe atoms. The solid line has been obtained by
smoothing the data.} \label{fig:Asym_en_dep}
\end{figure}

Figure\ \ref{fig:Asym_en_dep} shows the measured values for the LRA
parameter as a function of the photoelectron energy for different
shells of Ar and Xe. In spite of the large error bars the data are
very consistent and the spread is small indicating that the
systematic errors were probably overestimated. The values show
definite structure with a zero crossing. This shape suggests a
general tendency for the asymmetry parameters. Few experimental
values measured for Xe $5p_{1/2}$ and $5p_{3/2}$ are outside of the
general trend. The reason of this deviation may be induced by
\hbox{$(4p)^{-1}$-$ns/md$} resonant excitations \cite{ricz2003}. The
photoelectron energy dependence of LRA parameter is approximately
the same for the $s$- and $p$-shells. However figure
\ref{fig:Asym_en_dep} shows that the zero crossing energies may be
different for the $s$- and $p$-shell photoelectrons. Further
experimental studies are necessary to understand the photoelectron
energy dependence of the LRA parameters for different atomic shells
with smaller errors. An interesting measurement would be to
investigate the energy dependence of LRA with different polarization
of the photon beam (left- and right-circularly polarized,
unpolarized beams).

\section{Conclusion}

The left-right asymmetry parameters were measured for
photoionization of the outer $s$-shells of noble gas atoms and H$_2$
molecule using linearly polarized synchrotron radiation
\cite{Ricz2007}. The aim of the present investigations were to
verify the exitance of the earlier found left-right asymmetry in the
angular distribution of photoelectrons. The recent measurements were
carried out in a totally different experimental environment
(applying another synchrotron light source and a different electron
spectrometer). The measured asymmetry parameters ($A_{LR}$)
resulting from two different studies are in good agreement and
significantly differ from zero. Both data sets show a decrease of
$A_{LR}$ with increasing atomic mass for $s$-shells. The
experimental LRA parameters of Ne 2$p$, Ar 3$p$, Kr 4$p$ and Xe 5$p$
were presented as well. The sign and the shape of the measured data
for $p$-shells differ from the data determined for $s$-shells. This
indicates that the unknown correlation is sensitive to the angular
momentum of the atomic shells in photon-atom interactions.

The data measured and published earlier were evaluated again for the
determination of the LRA parameters.  In the case of Ar 3$p$
photoelectrons the asymmetry parameters oscillate strongly in the
vicinity of the \hbox{$(2p)^{-1}$-$ns/md$} resonances. The
photoelectron energy dependence of the experimental LRA parameters
was also investigated for $s$- and $p$-shells of Ar and Xe atoms.

Recently the LRA was observed for Ar $2p$ shells by Heinasmaki {\it
et al} \cite{Heinasmaki2013} using two electrostatic analyzers at
the MAX\ II synchrotron. Their results agree well with our
experimental data. Trinter {\it et al} \cite{Trinter2012} have found
LRA for the angular distribution of the ejected photoelectrons and C
K-Auger electrons using left and right circularly polarized light
and carbonmonoxide molecular target.

The exclusion of the possible instrumental sources that could
produce the left-right asymmetry and the experimental results show
that the observed left-right asymmetry may be a real physical
process. Currently there is no physical explanation for a non-zero
asymmetry parameter which violates space inversion symmetry in the
photon-atom interaction. Theoretical models and further experiments
are required to understand the origin of the observed left-right
asymmetry.

\section{Acknowledgments}

We would like to give special thanks to Professor Seppo Aksela and his group in Oulu, Finland
 for participating in the development of the experimental setup and the scientific cooperation in the data
  collection at MAX\ II synchrotron in Lund, Sweden. We thank the members of the Atomic Physics
   Division of Atomki for useful discussions. The assistance of the staff of DORIS\ III at HASYLAB
   is gratefully acknowledged. This work was supported by the Hungarian Scientific Research Foundation (OTKA K104409)
    and the T\'AMOP-4.2.2.A-11/1/KONV-2012-0060 project \hbox {(cofinanced by EU and the European Social
    Fund).}

\section{Appendix}

 Here we discuss possible sources of the left-right asymmetry
 which might be due to experimental or instrumental inaccuracies.

\begin{flushleft}
(a) \emph{Mechanical inaccuracy}:
\end{flushleft}
   The shapes of the cylinder and the sphere of the analyzer may differ from the ideal
   form being
   slightly eccentric. According to the precise check of the shape the mechanical
   inaccuracy is less than 10 $\mu$m over the entire circumferences of the cylinder and the sphere,
    for both analyzers (ESA-22L and ESA-22G \cite{Ricz2007, Ricsoka2009}).
    It may cause less than $5$x$10^{-5}$ uncertainty in the
    left-right asymmetry parameter.
     However, these eccentricities cause the same intensity distortions for
     the Auger- and photoelectrons, therefore the normalization of
   the photoelectron intensity to an Auger electron intensity eliminates
    the influence of the mechanical uncertainty to the left-right asymmetry parameter.

\begin{flushleft}
   (b) \emph{Angular inaccuracy}:
\end{flushleft}
   An angular difference may exists between the real photon beam propagation direction and the beam axis
   defined by the spectrometer slit system. This angular misalignment is less than  $0.28^{\circ}$ in the present
   experiment and $0.19^{\circ}$ in the previous ones \cite{Ricz2007}.
    The values of the left-right asymmetry parameter caused by these misalignments are $5.9$x$10^{-7}$ and $3.2$x$10^{-7}$, respectively.
   Similar effect can be produced by the azimuthal rotation of the holder of the channeltrons.
   The possible largest rotation angle relative to the proper position is $0.05^{\circ}$ for both analyzers (ESA-22G and ESA-22L)  resulting not more than $4.9$x$10^{-8}$ left-right asymmetry.

\begin{flushleft}
   (c) \emph{Off-axis alignment of the spectrometer}:
\end{flushleft}
   When the axis of the spectrometer does not cross the axis of the photon
   beam (parallel shift), it may cause an acceptance angle difference between the two spectrometer halves.
 $0.5$\ mm deviation between the two axes may produce about $\Delta\psi=0.05^{\circ}$ and $\Delta\chi=0.02^{\circ}$
 differences in the acceptance angles for both analyzers.
   The value of the left-right asymmetry parameter caused by these differences is about 0.022.
   However, our standard normalization procedure
   (i.e. the intensity of the photoelectrons are normalized to the intensity of the Auger electrons)
   eliminates the effects of these inaccuracies because the solid angle is the same for Auger- and photoelectrons.

\begin{flushleft}
   (d) \emph{Magnetic field}:
\end{flushleft}
   Both spectrometers (ESA-22L and ESA-22G) are shielded by three layers of $\mu$-metal against the Earth magnetic field.
   The value of the residual magnetic field is less than $500$ nT in the scattering plane and in the analyzer.
   Nevertheless, this weak magnetic field rotates slightly the trajectories, therefore may modify
   the original angular distributions of the electrons.
   The rotation angles of the electron trajectories are about $0.0053^{\circ}$ and $0.0037^{\circ}$ for 40 eV and 80 eV pass
   energies, respectively. If the electron momentum vector is perpendicular to the magnetic induction vector during
   the flight of the electrons, the values of the corresponding left-right asymmetry parameter are only
    $A_{LR}$=$4$x$10^{-9}$ and $A_{LR}$=$2.8$x$10^{-9}$, respectively. It is even lower than the value determined for the
    angular uncertainty discussed above.

\begin{flushleft}
   (e) \emph{Contamination of the electrodes}:
   \end{flushleft}
   Dirty surfaces of the analyzer electrodes may produce local charging which modifies the trajectory
   and the angular distribution of emitted electrons.
   This effect increases with decreasing pass energy.
   The pass energy dependence of the photoelectron angular distribution was
   checked experimentally at beam line I411 of the MAX\ II synchrotron using the ESA-22L electron spectrometer
   \cite{ricz2002}. The Ar 2$p$ photoelectron line was measured at 440 eV photon energy with different deceleration ratios ($E_{kin}$/$E_{pass}$) which varied between 1 and 9 (20-190 eV pass energy). The values of the left-right asymmetry parameters were constant within $20$~\% in this wide pass energy range.
      This investigation demonstrates the cleanness of the electrode surfaces
   and excludes any contribution of charging effect to the left-right
   asymmetry.
\newline
\begin{flushleft}
   (f) \emph{Exit slit of the monochromator}:
\end{flushleft}
Changing the exit slit size of the monochromator may modify the
dimension and the
   intensity distribution of the photon beam in the target region. This may produce
   a "misalignment" effect similar to the geometrical misalignments discussed above.
   These investigations were carried out at such photon energy
    where the photon flux had maximum value. The detected count rates of the channeltrons were
    also maximum in these measurements relative to the "normal" measurements. The count rates of
    every channeltron varied with a factor of 100 between
    the minimum and maximum values of the monochromator exit slit size.
    It was found that the value of the left-right asymmetry parameter does not depend on the slit size of the monochromator within $0.4$~\% and  $2.5$~\% error in the present and earlier experiments, respectively.
    This independence also indicates that the signal processing and
    the speed of the scalers were sufficiently fast and dead time effects were negligible.

\end{document}